\documentclass[a4paper, amsfonts, amssymb, amsmath, reprint, showkeys, nofootinbib, twoside]{revtex4-1}
\usepackage[english]{babel}
\usepackage[utf8]{inputenc}
\usepackage[colorinlistoftodos, color=green!40, prependcaption]{todonotes}

\usepackage{hyperref}
\usepackage{amsmath}

\usepackage{mathtools}
\usepackage{physics}
\usepackage{xcolor}
\usepackage{graphicx}
\usepackage[left=23mm,right=13mm,top=35mm,columnsep=15pt]{geometry} 
\usepackage{adjustbox}
\usepackage{placeins}
\usepackage[T1]{fontenc}
\usepackage{lipsum}
\usepackage{csquotes}

\usepackage{amsthm}
\usepackage{mathtools}
\usepackage{physics}
\usepackage{xcolor}
\usepackage{graphicx}
\usepackage[left=23mm,right=13mm,top=35mm,columnsep=15pt]{geometry} 
\usepackage{adjustbox}
\usepackage{placeins}
\usepackage[T1]{fontenc}
\usepackage{lipsum}
\usepackage{csquotes}

\usepackage{color}

\usepackage{subcaption}
\begin{document}

\title{Fundamental limits of parameter estimation with heralded optical non-Gaussian states generated from Gaussian resources}

\author{Shohei Kiryu$^{1}$}
\email{kiryu.opt@keio.jp}
\author{Kazufumi Tanji$^{1}$}
\author{Yoshihiro Ueda$^{1}$ }
\author{Kosuke Fukui$^{2}$ }
\author{Masahiro Takeoka$^{1,3}$}

\affiliation{$^{1}$ Department of Electronics and Electrical Engineering, Keio University, 3-14-1 Hiyoshi, Kohoku-ku, Yokohama 223-8522, Japan\\
$^{2}$ Department of Applied Physics, School of Engineering, The University of Tokyo, 7-3-1 Hongo, Bunkyo-ku, Tokyo 113-8656, Japan\\
$^{3}$ National Institute of Information and Communications Technology (NICT), Koganei, Tokyo 184-8795, Japan}

\date{\today}

\begin{abstract}
    Non-Gaussian states can exhibit large quantum Fisher information (QFI) in quantum sensing. In optical systems, however, its generation is often probabilistic via the boson-sampling type conditional operation and thus its generation rate is limited. This probabilistic generation of non-Gaussian resource should be taken into account for evaluation of the sensing performance. Then a natural question arising is whether the use of heralded probabilistic non-Gaussian states is better than that of the original deterministic Gaussian states for quantum sensing. 
    In this paper, we answer to this question for single-parameter phase-estimation.  
    By using photon-number conservation in passive linear optical systems, we show that heralded state preparation before parameter encoding can be mapped to a postselection problem after parameter encoding for phase estimation. This mapping allows the success probability of heralding to be included naturally in the metrological performance. 
    We introduce an effective quantum Fisher information (EQFI), defined as the success-probability-weighted QFI of the heralded outputs, and prove that it cannot exceed the QFI of the original Gaussian inputs. The result highlights the importance of resource counting in quantum sensing toward better understanding of the resource efficient advantage of optical quantum sensing. 


\end{abstract}

\maketitle


\section{INTRODUCTION}
   Photonic architectures present unique advantages for quantum technologies. These systems maintain quantum coherence effectively at room temperature. They also naturally support communication over long distances~\cite{takeda2019toward}.
   
   Photonic quantum states are classified as Gaussian or non-Gaussian states. Nowadays, generation of Gaussian states is highly matured~\cite{yoshikawa2007demonstration,ukai2011demonstration,su2013gate,kashiwazaki2021fabrication,larsen2021deterministic,nehra2022few,hirota2025generation,yoshida2025sequential}, while the deterministic generation of non-Gaussian states remains a significant challenge. 
   Since non-Gaussian resources are required for many continuous-variable (CV) quantum information protocols~\cite{lloyd1999quantum,Cochrane1999Cat,Gottesman2001GKP,bartlett2002efficient,Jeong2002Cat,Ralph2003Cat,ralph2005loss,menicucci2006universal,varnava2006loss,hilaire2023linear,reiss2026optimal}, 
   generating and controlling non-Gaussian states therefore constitutes a primary objective for CV architectures.
   
   Recent experiments have employed a form of Gaussian boson sampling~\cite{hamilton2017gaussian} for the probabilistic generation of complex non-Gaussian states~\cite{konno2024logical,endo2025high,larsen2025integrated,yu2026extensible}, including the cat states~\cite{Jeong2002Cat} and the Gottesman–Kitaev–Preskill states~\cite{Gottesman2001GKP}. 
   This approach consists of interfering Gaussian states and subsequent photon-number-resolving measurements in the partial output ports. The non-Gaussian state is generated when the specific photon number $\mathbf{n}_B$ is detected. This approach is highly versatile because different photon-number detection patterns allow the generation of a wide range of non-Gaussian states. 
   It nevertheless encounters a severe scaling limitation since the success probability decreases exponentially with respect to the system size~\cite{dakna1997generating,lund2004conditional,takase2023gottesman,aghaee2025scaling,solodovnikova2025loss,hanamura2025beyond}.

   Non-Gaussian optical states are also of significant interest as probe states for quantum sensing. In parameter estimation tasks, the quantum Fisher information (QFI) provides a fundamental figure of merit that determines the attainable precision~\cite{giovannetti2011advances}. In this context, conditionally generated non-Gaussian states can exhibit large QFI and have been explored in various quantum-metrological settings~\cite{nagata2007beating,huver2008entangled,joo2011quantum,israel2012experimental,daryanoosh2018experimental,gessner2019metrological,tatsuta2019quantum,oh2020optical,hanamura2021estimation,zhou2025phase,park2025quantum}. However, heralded generation introduces a substantial overhead due to its small success probability and the need for substantial Gaussian input resources. This overhead leads to a trade-off between the conditional metrological gain and the state generation rate, which raises a natural question: whether use of the heralded non-Gaussian states is beneficial for quantum sensing?  

    Previous studies have shown that postselection applied to parameter-encoded states cannot improve the average metrological precision once the success probability and the information contained in discarded events are properly taken into account~\cite{combes2014quantum,zeng2025non}. However, this no-go theorem is not directly applicable to our question since the heralding by conditional measurement is completed before the unknown parameter is imprinted on the probe state.

    
    In this work, we answer to the above question by an alternative no-go theorem. We establish a fundamental resource accounting limit for phase estimation with heralded non-Gaussian states generated from Gaussian inputs. 
    We introduce an effective quantum Fisher information (EQFI), which quantifies the average information yield per use of the input Gaussian resource and enables a resource-fair comparison between probabilistic non-Gaussian probes and deterministic Gaussian probes. We prove that this EQFI is upper bounded by the QFI of the original Gaussian input state. 
    That is, use of the heralded non-Gaussian states cannot give a net metrological advantage once the full input resource cost is taken into account.
    
    \begin{figure*}[t] 
        \centering  \includegraphics[width=0.99\textwidth]{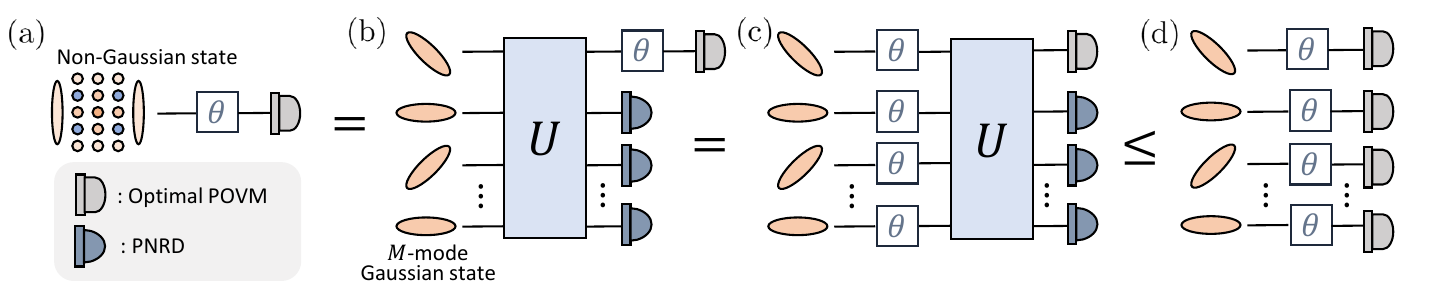}
        \caption{(a) Quantum sensing with a non-Gaussian state and (b) heralded non-Gaussian states. (c) Equivalent circuit with (b) with respect to the QFI. (d) Quantum sensing with Gaussian states. $U$ is a linear optical unitary operation. 
        }
        \label{FIG1_GraphicalAbstract}
    \end{figure*} 
    
\section{GAUSSIAN AND NON-GAUSSIAN RESOURCES}
    Consider a system of $M$ modes. A separable pure Gaussian state $\ket{\Psi_G}$ is prepared by applying local Gaussian unitary operations to the vacuum state $\ket{0}^{\otimes M}$, which can be written as
    \begin{equation}
        \ket{\Psi_G} = \bigotimes_{k=1}^{M} \hat{G}_k \ket{0}_k.
    \end{equation}
    Here, the Gaussian unitary operator $\hat{G}_k$ acting on the $k$-th mode is defined using the displacement operator $\hat{D}_k$ and the squeezing operator $\hat{S}_k$ as
    \begin{equation}
        \hat{G}_k = \hat{D}_k(\alpha_k)\hat{S}_k(\xi_k).
    \end{equation}
    The displacement operator $\hat{D}_k$ is given by
    \begin{equation*}
    \hat{D}_k(\alpha_k) = \exp\left(\alpha_k \hat{a}_k^\dagger - \alpha_k^* \hat{a}_k\right),
    \end{equation*}
    where $\alpha_k$ is the complex displacement parameter. The squeezing operator $\hat{S}_k$ is expressed as
    \begin{equation*}
        \hat{S}_k(\xi_k) = \exp\left[\frac{1}{2}\left(\xi_k^* \hat{a}_k^2 - \xi_k (\hat{a}_k^\dagger)^2\right)\right],
    \end{equation*}
    where $\xi_k = r_k e^{i\theta_k}$ is the complex squeezing parameter. In these definitions, $\hat{a}_k^\dagger$ and $\hat{a}_k$ denote the bosonic creation and annihilation operators for the $k$-th mode, respectively.
    A linear optical unitary transformation $\hat{U}$ is applied to the $M$-mode system. The modes are then partitioned into an output subsystem $A$ and an ancillary subsystem $B$ consisting of $j$ modes. Photon-number-resolving detection (PNRD) performed on subsystem $B$ yields a photon-number pattern $\mathbf{n}_B = \{n_k\}_{k \in B}$.
    The projection onto the measurement outcome $\mathbf{n}_B$ yields the conditional pure state $\ket{\psi(\mathbf{n}_B)}_{\text{out}}$ in the output subsystem $A$ given by
    \begin{equation}\label{Eq:HeraldedState}
        \ket{\psi(\mathbf{n}_B)}_{\text{out}} = \mathcal{N} \bra{\mathbf{n}_B} \hat{U} \ket{\Psi_G},
    \end{equation}
    where $\bra{\mathbf{n}_B}_B = \bigotimes_{k \in B} \bra{n_k}_k$ represents the detected multi-mode Fock state, and $\mathcal{N}$ is a normalization constant. This scheme, known as heralded generation, allows to create non-Gaussian states from Gaussian resources.

\section{LIMIT OF PHASE ESTIMATION VIA HERALDED NON-GAUSSIAN STATES}

    In this section, we consider the phase estimation by optical non-Gaussian probe states (Fig.~\ref{FIG1_GraphicalAbstract}(a)) where $\theta$ denotes the unknown phase shift to be estimated. As illustrated in Fig.~\ref{FIG1_GraphicalAbstract}(b), we assume that the non-Gaussian state is generated from the Gaussian boson sampling setup where $M$-mode separable pure Gaussian states are processed by a linear optical circuit and then partially measured by PNRDs. Depending on the outcome of PNRDs, a nontrivial non-Gaussian state is conditionally generated (heralded) with some success probability. In practice, this conditional state generation process should be properly included in the evaluation of the phase estimation performance. 
    In the following, we show that the parameter estimation performance of this heralded non-Gaussian strategy cannot be better than that of the Gaussian phase estimation strategy where the original $M$-mode Gaussian states are directly used as the probe states (Fig.~\ref{FIG1_GraphicalAbstract}(d)).

    The phase estimation performance is evaluated by the quantum and classical Fisher information formalisms. 
    Suppose the unknown parameter $\theta$ is encoded in $|\psi(\theta)\rangle$. 
    The state is detected by some measurement and suppose the probability distribution of the measurement outcome is $\{ P_i \}_i$. Then the variance of the estimator $\Delta \theta^2$ is bound by the Cramér-Rao bound, 
    \begin{equation}\label{Eq:CRB}
    \Delta \theta^2 \ge \frac{1}{n F_\theta},  
    \end{equation}
    where $n$ is the number of trials and 
    \begin{equation}\label{Eq:CFI}
    F_\theta = \sum_i \frac{1}{P_i} \left(\frac{\partial P_i}{\partial \theta}\right)^2,  
    \end{equation}
    is the classical Fisher information (CFI). 
    It is known that the CFI is upper bounded by the quantum Fisher information (QFI) which is solely a function of the quantum state. For a pure input state and phase estimation, the quantum Fisher information (QFI) is known to be   
    \begin{equation}\label{Eq:QFI}
    F_Q = 4(\Delta \hat{N})^2 = 4 \left( \bra{\psi} \hat{N}^2 \ket{\psi} - \bra{\psi} \hat{N} \ket{\psi}^2 \right), 
    \end{equation}
    where $\hat{N}$ is the number operator. 
    For a single-parameter estimation, it is known that there exists a non-collective measurement that saturates the QFI.

    Let us start by formulating the heralded non-Gaussian states in Fig.~\ref{FIG1_GraphicalAbstract}(b). Partition the $M$ modes into subsystems $A$ and $B$, for the non-Gaussian output mode and the ancillary modes to be detected by PNRDs, respectively. 
    In the figure, $U$ denotes the unitary operator representing the passive linear optical circuit, and let $\mathbf{n}_B=\{n_k\}_{k\in B}$ be a set of the PNRD outcomes, i.e. the photon number pattern detected in subsystem $B$.
    When the PNRD observes $\mathbf{n}_B$, the conditional non-Gaussian state in subsystem $A$ is given by Eq.~\eqref{Eq:HeraldedState}. Throughout this paper, we assume that subsystem $A$ is a single-mode system. However, its generalization to a multi-mode system is straightforward. 
    Denote the phase operator for the unknown phase shift as $\hat{R}_A(\theta) \equiv e^{i \theta \hat{n}_A}$. Applying this phase shift into the conditional non-Gaussian state, we get 
    \begin{equation}\label{Eq:NonGaussianWithPhase}
    \begin{split}
        \ket{\psi(\theta, \mathbf{n}_B)}_{\rm out} &= \hat{R}_A(\theta) \ket{\psi(\mathbf{n}_B)}_{\rm out} \\
        &= \mathcal{N}_{\mathbf{n}_B} \hat{R}_A(\theta) \bra{\mathbf{n}_B}\hat{U}\ket{\Psi_G}. 
    \end{split}
    \end{equation}
    
    Usually, a desired non-Gaussian state is obtained only when a specific measurement outcome $\mathbf{n}_B^*$ is obtained. 
    While use of this state could give better quantum or classical Fisher information than that of Gaussian states, its generation is probabilistic and in practice, the success probability of state preparation should be taken into account for comparison.  
    Consider the optimal measurement for given $\ket{\psi(\theta, \mathbf{n^*}_B)}_{\rm out}$ to maximize its CFI, and let $x$, $P(x|\mathbf{n}_B^*, \theta)$, and $F(X|\mathbf{n}_B^*)$ be the measurement outcome, its probability distribution, and the corresponding CFI. 
    
    In the heralding scenario, $\mathbf{n}_B^*$ is conditionally obtained with probability $P(\mathbf{n}_B^*)$. 
    This means that only $n P(\mathbf{n}_B^*)$ of $n$ trials can be used for the phase estimation. 
    Then the effective CFI applied to the Cramér-Rao bound with $n$ trials should be given by $ P(\mathbf{n}_B^*) F(X|\mathbf{n}_B^*)$

    The above scenario can be improved when one uses not only specific PNRD outcome but all possible PNRD outcomes. Suppose, for each given $\mathbf{n}_B$, the corresponding optimal measurement is applicable. Then the effective CFI in this case is improved to be the average of all possible conditional states, i.e. for Fig.~\ref{FIG1_GraphicalAbstract}(b), we have 
    \begin{equation}\label{Eq:F^(b)}
    \begin{split}
    F_\theta^{({\rm b})} &= \sum_{\mathbf{n}_B} P(\mathbf{n}_B) F(X|\mathbf{n}_B)\\
    & = \sum_{\mathbf{n}_B} P(\mathbf{n}_B) F_Q\left( \ket{\psi(\theta, \mathbf{n}_B)}_{\rm out}  \right) \\
    & \equiv \bar{F}_Q,
    \end{split}
    \end{equation}
    where the second equality follows from the fact that for the single-parameter estimation, the optimal measurement attaining the QFI always exists. 
    We call $\bar{F}_Q$ the effective QFI (EQFI). 

    Now, let us slightly modify the strategy in Fig.~\ref{FIG1_GraphicalAbstract}(b). Instead of considering each heralded non-Gaussian states, consider the state before PNRDs, i.e. $\hat{U}|\Psi_G\rangle$ as the probe state and include PNRDs in a part of the measurement for the phase estimation. 
    Then the probability distribution for the phase estimation is given by the joint probability 
    \begin{equation}
        P(x,\mathbf{n}_B|\theta)=P(\mathbf{n}_B)P(x|\mathbf{n}_B,\theta), 
    \end{equation}
    and its CFI is given by $F_\theta(X, \mathbf{n}_B)$. 
    Then we observe  
    \begin{equation}\label{Eq:(b)(b')}
    \begin{split}
        F_\theta^{({\rm b})} &=  
        F_\theta(\mathbf{n}_B)+\sum_{\mathbf{n}_B}P(\mathbf{n}_B|\theta)F_\theta(X|\mathbf{n}_B)
         \\
        &= F_\theta(X, \mathbf{n}_B) \equiv F_\theta^{({\rm b'})}, 
    \end{split}
    \end{equation}
    where the first equality follows from $F_\theta(\mathbf{n}_B)=0$, which is guaranteed by the fact that $P(\mathbf{n}_B)$ is independent of $\theta$. 
    The second equality follows from the chain rule of the CFI. 
    Here we insert the phase operator
    \begin{equation}\label{Eq:R_B}
         \hat{R}_B(\theta) = e^{i\theta \sum_k \hat{n}_k^B}, 
    \end{equation}
    in front of the PNRDs in subsystem $B$, 
    where $\hat{n}_k^B$ is the number operator for the $K$-mode in subsystem $B$. 
    Since 
    \begin{equation}
        \bra{\mathbf{n}_B}\hat{R}_B(\theta) = e^{i\theta N_B}\bra{\mathbf{n}_B}, 
    \end{equation}
    where $N_B$ is the total photon number detected by PNRDs in system B, the conditional state is 
     \begin{equation}\label{eq:approximate_PSIout}
    \begin{split}
        &\mathcal{N}_{\mathbf{n}_B}\bra{\mathbf{n}_B} 
        \left(
        \hat{R}_A(\theta)\otimes\hat{R}_B(\theta)
        \right)
        \hat{U}\ket{\Psi_G} \\
        &= e^{i\theta N_B}\mathcal{N}_{\mathbf{n}_B}\hat{R}_A(\theta)\bra{\mathbf{n}_B}\hat{U}\ket{\Psi_G}\\ 
        &= \ket{\psi(\theta, \mathbf{n}_B)}_{\rm out},
    \end{split}
    \end{equation}   
    where in the last equality, we ignore the global phase. 
    That is, the state does not change by adding the phase shifts in subsystem $B$ up to the global phase. This implies that its optimal measurement also does not change and the corresponding CFI is exactly equal to $F_\theta^{({\rm b'})}$. 

    An implication of the above extra phase term is that now we can commute the phase shifts and the passive linear optical circuit $U$. 
    To do so, we use photon-number conservation in passive linear optics.
    Since the passive linear optical circuit conserves the total photon number, we have
    \begin{equation} \label{eq:communication relation}
        [\hat{U},\hat{N}] = 0,
    \end{equation}
    where $\hat{N}=\sum_{i=1}^M\hat{n}_i$ which is the total number operator for the $M$ modes. 
    This implies that $\hat{U}$ also commutes with the phase rotations $\hat{R}_{\rm total}(\theta)=\hat{R}_A(\theta) \otimes \hat{R}_B(\theta) =e^{i\theta \hat{N}}$, and hence
    \begin{equation}
        [\hat{U},\hat{R}_{\rm total}(\theta)] = 0.
    \end{equation}
    Equivalently, this can be written as
    \begin{equation}
        \hat{R}_{\rm total}(\theta)\hat{U} = \hat{U}\hat{R}_{\rm total}(\theta).
    \end{equation}
    A detailed proof of (\ref{eq:communication relation}) is given in Appendix~\ref{Appeix:CommutationRelation}.
    This means that applying the phase rotation after the linear optical circuit gives the same result as applying it before the circuit.
    By swapping the phase shifts and $U$ in Fig.~\ref{FIG1_GraphicalAbstract}(b), we obtain Fig.~\ref{FIG1_GraphicalAbstract}(c). Since this does not change the measurement probability distribution, we clearly have 
    \begin{equation}\label{Eq:(b')(c)}
        F_\theta^{({\rm b'})} = F_\theta^{({\rm c})} , 
    \end{equation}    
    where $F_\theta^{({\rm c})}$ is the CFI for Fig.~\ref{FIG1_GraphicalAbstract}(c).

    In Fig.~\ref{FIG1_GraphicalAbstract}(c), $U$ and all measurements are regarded as a collective measurement, which is not necessarily optimal to estimate $\theta$. 
    Consider to replace it with the optimal measurement. Since the Gaussian input is an $M$-mode separable states, the optimal measurement attaining the QFI exists and is given by the separable measurement. Then we have 
    \begin{equation}\label{Eq:(c)(d)}
        F_\theta^{({\rm c})} \le F_Q\left(\hat{R}_{\rm total}(\theta)\ket{\Psi_G}\right) \equiv F_Q^{\rm in}. 
    \end{equation}    
    Combining Eqs.~(\ref{Eq:F^(b)}), (\ref{Eq:(b)(b')}), (\ref{Eq:(b')(c)}), and (\ref{Eq:(c)(d)}), we have 
    \begin{equation}\label{Eq:result_bound}
        \bar{F}_Q^{\rm out} \le F_Q^{\rm in}. 
    \end{equation}    
    This conludes that the EQFI for conditional non-Gaussian states can never be better than the QFI of the original Gaussian input states. 
    Since this is a single-parameter estimation, the separable measurement attaining $F_Q^{\rm in }$ always exists (Fig.~\ref{FIG1_GraphicalAbstract}(d)).  
    Note that this statement holds for any Gaussian input states.

     The right-hand side of (\ref{Eq:result_bound}) is further bounded by the optimal input states. Consider the optimal Gaussian inputs under the average input power constraint. 
     A pure separable $M$-mode Gaussian input state is characterized by a set of parameters, $\{ \alpha_k, r_k \}_{k=1,\cdots, M}$, where $\alpha_k$ and $r_k$ are displacement and squeezing for the $k$-th mode, respectively.  
     According to \cite{oh2019optimal}, the QFI for a Gaussian sate in the $k$-th mode is given by  
    \begin{equation}
    \begin{aligned} &F_{\text{Gauss}}(\alpha_k, r_k) \\ &= \begin{cases} 4e^{2r_k}|\alpha_k|^2 \\ \quad (\text{if } |\alpha_k| \ge \sqrt{2} e^{-r_k}\sinh 2r_k), \\ \frac{1}{2}\left[ 2\sinh 2r_k + (1+\coth 2r_k)|\alpha_k|^2 \right]^2 \\ \quad (\text{if } |\alpha_k| < \sqrt{2} e^{-r_k}\sinh 2r_k), \end{cases} \end{aligned} \end{equation}
    where its average power is given by $\bar{n}_k = \sinh^2 r_k + |\alpha_k|^2$. 
    Under the power constraint, the QFI is maximized when $\bar{n}_k = \sinh^2 r_k$, in other words, when the input is a squeezed vacuum. Then its QFI has a simple form,   
    \begin{equation}\label{Eq:F_Gauss^opt}
        F_{\rm Gauss}^{{\rm opt} (k)} = 8 \bar{n}_k (\bar{n}_k +1). 
    \end{equation}
    In this case, it is known that the above QFI is attained by standard homodyne detection~\cite{monras2006optimal,oh2019optimal,fadel2025quantum}. 
    Finally, optimized the power allocation among $M$ modes. Let $\bar{n}=\sum_k \bar{n}_k$. 
    Since Eqs.~(\ref{Eq:F_Gauss^opt}) is a convex function, the optimal power allocation is $\bar{n}_1=\bar{n}$ and $\bar{n}_k=0$ for $k=2,\cdots,M$. Thus the optimal QFI for an $M$-mode Gaussian input is 
    \begin{equation}\label{Eq:F_Gauss^opt2}
        F_{\rm Gauss}^{{\rm opt}} = 8 \bar{n} (\bar{n} +1),  
    \end{equation}
    which is simply attained by standard homodyne detection. 

\if
     that for practically important cases, 
     Based on the optimal Gaussian metrology \cite{oh2019optimal}, we characterize the maximum Fisher information $F_{\text{Gauss}}(\alpha_k, r_k)$ for each independent mode $k$. The resulting expression is piecewise. It takes different forms depending on whether the system is dominated by the displacement $\alpha_k$ or the squeezing $r_k$ of the input states. We assume optimal phase matching to rigorously determine this function. The resulting expression is
    \begin{equation}
    \begin{aligned} &F_{\text{Gauss}}(\alpha_k, r_k) \\ &= \begin{cases} 4e^{2r_k}|\alpha_k|^2 \\ \quad (\text{if } |\alpha_k| \ge \sqrt{2} e^{-r_k}\sinh 2r_k), \\ \frac{1}{2}\left[ 2\sinh 2r_k + (1+\coth 2r_k)|\alpha_k|^2 \right]^2 \\ \quad (\text{if } |\alpha_k| < \sqrt{2} e^{-r_k}\sinh 2r_k). \end{cases} \end{aligned} \end{equation}
    Each of the $M$ modes contributes independently to the overall precision. We define the total input precision limit under Gaussian operations by summing these terms to obtain $F^{\text{in}}_{\text{Gauss}} = \sum_{k=1}^M F_{\text{Gauss}}(\alpha_k, r_k)$. 
        
    We now restrict our analysis to specific initial conditions. These encompass pure squeezed vacuum states with $\alpha_k=0$ and coherent states with $r_k=0$. Applying our earlier findings to the established inequality yields a fully operational bound. This rigorous relation is expressed as
    \begin{equation}
    F^{\text{in}}_{\text{Gauss}} \ge \bar{F}^{\text{out}}_\theta.
    \end{equation}
    Standard homodyne detection strictly saturates the ultimate quantum Fisher information for these fundamental Gaussian resources~\cite{monras2006optimal,oh2019optimal,fadel2025quantum}. Our theoretical boundary therefore translates directly into a practical threshold. This conclusion highlights a profound physical reality regarding quantum measurement strategies. Standard homodyne measurements on deterministic squeezed vacuum or coherent states provide a substantial advantage. The average precision obtained from these initial states intrinsically matches or exceeds the precision achieved through far more complex procedures.
    When each input mode is either a squeezed vacuum state or a coherent state, an appropriate homodyne measurement saturates the QFI of the input Gaussian state. Therefore, the optimal measurement allowed in Fig.~\ref{FIG1_GraphicalAbstract}(c) and the standard Gaussian measurement shown in Fig.~\ref{FIG1_GraphicalAbstract}(d) yield the same attainable Fisher information.
\fi

    \begin{figure*}[t] 
        \centering  \includegraphics[width=0.99\textwidth]{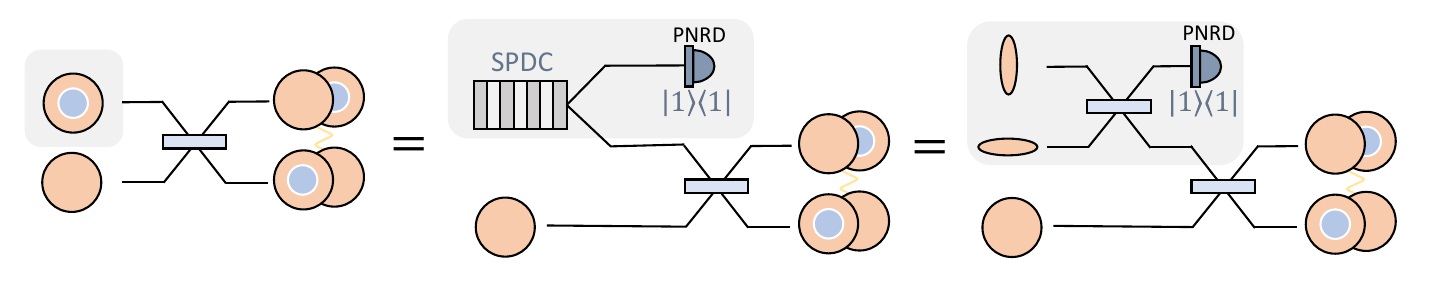} \caption{Optical setup to generate single-photon entangled states. The input single photons are heralded from two-mode squeezed vacuum states produced via spontaneous parametric down-conversion. The entire optical circuit can also be decomposed into an equivalent interferometric system. This system consists of two squeezed vacuum states and a half beam splitter.} \label{FIG2_PhotonEqualy}
    \end{figure*} 
    
\section{Example: Heralded single photon RESOURCES}
    In optical quantum information processing, photon-number states such as the vacuum state $\ket{0}$ and the single-photon state $\ket{1}$ serve as fundamental resources. The deterministic generation of ideal single photons remains a practical challenge. This challenge has motivated alternative approaches based on weak coherent light or heralded sources of single photons, which rely on spontaneous parametric down-conversion (SPDC). This process generates a two mode squeezed vacuum state, and subsequent photon detection in the idler mode maps the signal mode to the single-photon state. This photon number state generation based on SPDC is represented by the following equivalent procedure. First, the state generated by SPDC can be represented by an equivalent circuit involving squeezed vacuum states and a half beam splitter described by
    \begin{equation}
        B_{12}\hat{S}_1(\xi)\hat{S}_2(-\xi)\ket{0}_1\ket{0}_2 = \sqrt{1-\lambda^2}\sum_{n=0}^\infty \lambda^n\ket{n}_1\ket{n}_2 ,
    \end{equation}
    where $\lambda = \tanh(r)$. 
    We then perform a photon number measurement on the first (idler) mode. If the measurement outcome is one photon, the projection onto the first mode prepares the second (signal) mode in the single-photon state. This demonstrates an equivalence to the transformation illustrated in the Fig. \ref{FIG2_PhotonEqualy}.
    
    We now consider generating a path entanglement of the photon number states using this source. A half beam splitter receives the generated single photon $\ket{1}$ to produce an entangled state. We express this resulting state as
    \begin{equation}
        \label{eq:single-photon_path_entanglement}
        B_{23}\ket{1}_2\ket{0}_3 = \frac{1}{\sqrt{2}} \left( \ket{1}_2\ket{0}_3 + \ket{0}_2\ket{1}_3 \right).
    \end{equation}
    Combining the single photon source and the generation of the path entanglement, the input is formed by two squeezed vacuum states with different phases and a vacuum state. The measurement process prepares the targeted entangled state of single photons. In the following parts, we compare the input state with the resulting path entanglement based QFI and EQFI.

    We first evaluate the QFI of the raw SPDC output. The exact sum for QFIs from the two initial squeezed vacuum states takes the form
    \begin{equation}\label{Eq:SqueezedVacuumQFI}
        2F_Q (\ket{\xi}) = 16(\sinh^2 r + 1)\sinh^2 r.
    \end{equation}
    Next, we consider the QFI of the heralded entanglement state. While we consider the single-photon path entanglement in Eq.~\eqref{eq:single-photon_path_entanglement}, we generalize it to $n$-photon path entanglement below. If $n$ photons are detected in the first mode of the SPDC, the $n$-photon state $\ket{n}$ is generated in the second mode. We inject this state into a beam splitter. The process yields an entangled state of $n$ photons:
    \begin{equation}
        \frac{1}{\sqrt{n!}}\left( \frac{1}{\sqrt{2}}(\hat{a}^{\dagger}_1 + \hat{a}^{\dagger}_2) \right)^n \ket{0}.
    \end{equation}
    We then evaluate the QFI of this $n$-photon entangled states. Since these $n$ photon pairs are generated according to a probability distribution of the SPDC process, we evaluate the EQFI for a heralded source. Let $P(n)$ be the probability of detecting $n$ photons in the first mode. This probability equals $(1-\lambda^2)\lambda^{2n}$. The corresponding average information $\bar{F}_{\theta}$ becomes
    \begin{equation}
        \label{eq:CFI_TMSV}
        \bar{F}_{\theta} = \sum_{n=0}^{\infty} P(n) \cdot n = (1-\lambda^2)\sum_{n=0}^{\infty} n \lambda^{2n} = \sinh^2 r.
    \end{equation}
    The last transformation directly follows from the geometric series:
    \begin{equation}
    \sum_{n=0}^{\infty} n x^n = x \frac{d}{dx} \sum_{n=0}^{\infty} x^n = x \frac{d}{dx} \left( \frac{1}{1-x} \right) = \frac{x}{(1-x)^2}.
    \end{equation}
    We substitute this result into Eq.\eqref{eq:CFI_TMSV}, yielding
    \begin{equation}
        \bar{F}_{\theta} = (1-x) \cdot \frac{x}{(1-x)^2} = \frac{x}{1-x}.
    \end{equation}
    Restoring the original variable $x = \tanh^2 r$, we obtain the above form:
    \begin{equation}
        \bar{F}_{\theta} = \frac{\tanh^2 r}{1 - \tanh^2 r} = \frac{\sinh^2 r / \cosh^2 r}{1 / \cosh^2 r} = \sinh^2 r.
    \end{equation}
    The EQFI of the generated entangled state exactly matches the average SPDC photon number as $\bar{n} = \sinh^2 r$.
    This indicates that the measurement induces a projection operation and eliminates the natural quantum coherence of the squeezed state. The system therefore loses the unique advantage of photon number fluctuations. Consequently, the resulting sensitivity abruptly reduces to the fundamental shot noise limit. 
    In contrast, as shown in Eq.~\eqref{Eq:SqueezedVacuumQFI}, the original squeezed vacuum state possesses a QFI that scales as the order of $\sinh^4 r$, suggesting that the state prior to projection holds significantly higher potential as a metrological resource than heralded states.

\section{CONCLUSION}
    In this work, we evaluated the metrological utility of heralded non-Gaussian quantum resources generated from Gaussian inputs, explicitly taking into account the success probability of state generation. While previous studies have shown that postselection after parameter encoding cannot improve the average metrological precision when all outcomes are properly accounted for, heralded state preparation before parameter encoding is not directly covered by this conventional postselection no-go argument. We showed that, for phase estimation with passive linear optics, photon-number conservation maps this preparation-stage heralding to an equivalent postselection process after parameter encoding. Based on this equivalence, we introduced an effective quantum Fisher information (EQFI) that incorporates the generation probability, and proved that the average metrological information of the heralded output cannot exceed the intrinsic metrological potential of the input Gaussian state.

    The physical origin of this limitation is the conservation of the total photon number. Since passive linear optical transformations commute with the total photon-number operator, a phase rotation on all modes can be moved across the linear optical circuit. Moreover, the phase rotation acting on the measured modes contributes only a global phase after projection onto photon-number eigenstates. Thus, a large QFI observed in a conditional successful state does not represent an amplification of information, but rather a redistribution of the information already present in the input state into rare successful events. As a concrete example, we analyzed heralded photon-number states generated by SPDC. Although the successfully generated photon-number and path-entangled states can exhibit large conditional QFI, averaging over the photon-number detection probabilities yields $\bar{F}_\theta=\sinh^2 r$, showing that the high sensitivity scaling associated with the photon-number fluctuations of the original squeezed vacuum is lost.
    
    Our results do not rule out the usefulness of non-Gaussian resources or postselection in more broader quantum technologies. Non-Gaussian resources remain essential for quantum computation, quantum error correction, and quantum state engineering. The limitation identified here applies specifically to linear phase estimation with passive linear optics when heralded probe states are evaluated together with their generation probabilities. Under nonlinear parameter encoding, measurement constraints, detector limitations, or other specific experimental conditions, probabilistic state generation may still provide practical advantages. Within this regime, the bound is not restricted to a particular encoding of the output state, but applies to photonic continuous-variable (CV), discrete-variable (DV)~\cite{zavatta2004tomographic,mosley2008heralded,kaneda2015time,kaneda2016heralded}, and DV-CV hybrid bosonic resources~\cite{ andersen2015hybrid,lee2013near,jeong2014generation,lee2015nearly,omkar2020resource,omkar2021highly,lee2024fault,fukui2024resource,kiryu2025linear,bera2025long,kiryu2026linear} generated by passive linear optics and photon-number measurements.
    Our work therefore establishes a resource-accounting benchmark for assessing heralded non-Gaussian resources in quantum sensing.

\section*{Acknowledgements} 
    This work was supported by the Advancement of Next Generation Research Projects, Keio University; JST Moonshot R\&D Grant No.JPMJMS256E; JST ASPIRE Grant No.JPMJAP2427; JST SPRING Grant No.JPMJSP2123; JSPS KAKENHI Grant No.25K22795; JST PRESTO Grant No.JPMJPR23FA; and JST COI-NEXT Grant No.JPMJPF2221.
    
\bibliography{sample} 
\bibliographystyle{unsrt} 

\appendix
\section{Commutation relation between passive linear optics and global phase rotations}\label{Appeix:CommutationRelation}
    In this appendix, we provide a derivation of Eq.~\eqref{eq:communication relation}.

    According to the decomposition schemes of Reck et al.,~\cite{reck1994experimental} and Clements et al.,~\cite{clements2016optimal}, an arbitrary $M$-mode linear optical unitary operator $\hat{U}$ can be factorized as
    \begin{equation}
        \hat{U} = \hat{D} \prod_{(m,n) \in S} \hat{T}_{m,n}(\xi_{m,n}, \phi_{m,n}).
    \end{equation}
    Here, $S$ denotes the sequence of operations determined by the circuit topology. The operator $\hat{T}_{m,n}$ represents the fundamental unitary operation of a beam splitter and phase shifter acting on modes $m$ and $n$, acting as the identity operator on the remaining modes. Furthermore, $\hat{D}$ represents the product of single-mode phase shifts.
    
    We examine the commutation relations between these constituent elements and the total global phase shift operator $\hat{R}_{\text{total}}(\theta) = e^{i\theta\hat{N}}$. Here, $\hat{N} = \sum_{k=1}^M \hat{a}_k^\dagger \hat{a}_k$ is the total photon number operator.
    
    Linear optical elements such as beam splitters and phase shifters are passive operations and conserve the total photon number of the system. That is, the Hamiltonian for any fundamental operation $\hat{T}_{m,n}$ commutes with the total photon number operator $\hat{N}$ ($[\hat{N}, \hat{T}_{m,n}] = 0$). Due to this property, commutativity with the global phase operator is also established:
    \begin{equation}
    \begin{split}
        \hat{R}_{\text{total}}(\theta) \hat{T}_{m,n} &= e^{i\theta\hat{N}} \hat{T}_{m,n} \\
        &= \hat{T}_{m,n} e^{i\theta\hat{N}} \\
        &= \hat{T}_{m,n} \hat{R}_{\text{total}}(\theta).
    \end{split}
    \end{equation}
    By analogous reasoning, $[\hat{R}_{\text{total}}(\theta), \hat{D}] = 0$ also holds for $\hat{D}$.
    
    These individual commutation relations allow $\hat{R}_{\text{total}}(\theta)$ to freely propagate through the internal decomposition of the total unitary operator $\hat{U}$:
    \begin{equation}
    \begin{split}
        \hat{R}_{\text{total}}(\theta) \hat{U} &= e^{i\theta\hat{N}} \left( \hat{D} \prod_{(m,n) \in S} \hat{T}_{m,n} \right) \\
        &= \hat{D} e^{i\theta\hat{N}} \left( \prod_{(m,n) \in S} \hat{T}_{m,n} \right) \\
        &= \hat{D} \left( \prod_{(m,n) \in S} \hat{T}_{m,n} \right) e^{i\theta\hat{N}} \\
        &= \hat{U} \hat{R}_{\text{total}}(\theta).
    \end{split}
    \end{equation}
    
    This operator algebra rigorously derives the commutation relation $[\hat{R}_{\text{total}}(\theta), \hat{U}] = 0$. Physically, since the photon number is conserved within individual beam splitters and phase shifters, it means that applying a global phase operation either before or after the optical circuit yields the exact same result. Consequently, this establishes the identity $[\hat{U}, e^{i\theta\hat{N}}] = 0$ for arbitrary topologies and parameters.

\end{document}